\documentclass[aps,prd,superscriptaddress,onecolumn,showpacs,showkeys]{revtex4}
\usepackage{amssymb,amsmath,epsfig}
\usepackage{eurosym}
\usepackage{amsfonts}
\usepackage{array}
\usepackage{amsthm}
\usepackage{bm}
\usepackage{palatino}
\usepackage{mathpazo}
\usepackage{amssymb}
\usepackage{eurosym}
\usepackage{amsmath}
\usepackage{epsfig}
\usepackage{graphics}
\usepackage{color}
\usepackage{graphicx}
\usepackage{changes}
\usepackage{calrsfs}
\usepackage[colorlinks=true,
            linkcolor=red,
            urlcolor=blue,
            citecolor=blue]{hyperref}

\begin{document}

\title{\textbf{Effect of the hair on deflection angle by asymptotically flat black holes in Einstein-Maxwell-dilaton theory}}

\author{W. Javed} \email{wajiha.javed@ue.edu.pk;
wajihajaved84@yahoo.com}
\affiliation{Department of Mathematics, University of Education,\\
Township, Lahore-54590, Pakistan.}

\author{J. Abbas}
\email{jameelaabbas30@gmail.com}
\affiliation{Department of Mathematics, University of Education,\\
Township, Lahore-54590, Pakistan.}

\author{A. \"{O}vg\"{u}n}
\email{ali.ovgun@pucv.cl}
\homepage[]{https://www.aovgun.com}
\affiliation{Instituto de F\'{\i}sica, Pontificia Universidad Cat\'olica de Valpara\'{\i}%
so, Casilla 4950, Valpara\'{\i}so, Chile}
\affiliation{Physics Department, Faculty of Arts and Sciences, Eastern Mediterranean
University, Famagusta, North Cyprus, via Mersin 10, Turkey}

\begin{abstract}
In this paper, we are interested in a model of exact asymptotically
flat charged hairy black holes in the background of dilaton potential.
We study the weak gravitational lensing in the spacetime of hairy black
hole in Einstein-Maxwell theory with a non-minimally coupled dilaton and its
non-trivial potential. In doing so, we use the optical geometry of the flat
charged hairy black hole for some range of parameter $\gamma$. For this purpose,
by using Gauss-Bonnet theorem, we obtain the deflection angle of photon in a
spherically symmetric and asymptotically flat spacetime. Moreover, we also
investigate the impact of plasma medium on weak gravitational lensing by
asymptotically flat charged hairy black hole with a dilaton potential. Our analytically analyses show the effect of the hair on the deflection angle in weak field limits.
\end{abstract}
\date{\today}

\maketitle

\section{Introduction}

 At the darkest points in the universe, their boundaries perilous and invisible, space warps.
 The singularity constitutes the centre of a black hole and is hidden by the object's “surface,” the event horizon. A black hole is a location in space that possesses so much gravity, nothing can escape its pull, even light. It is said that fact is sometimes stranger than fiction, and nowhere is this more true than in the case of black holes. Since the first image of a black hole by Event Horizon Telescope \cite{Akiyama:2019cqa}, physicist now try to take even sharper images so that Einstein's theory of general relativity can be tested and also see the properties of the black holes \cite{R29,Konoplya:2019sns,Bambi:2019tjh,Shaikh:2019fpu,Abdikamalov:2019ztb,Abdujabbarov:2017pfw,Abdujabbarov:2016efm,Abdujabbarov:2016hnw,Atamurotov:2013sca} because there are many theoretically obtained black hole solutions with different properties.
 
According to Einstein, a black hole (BH) can be described only in terms of its mass and spin, 
which is known as "no-hair" theorem. No-hair means that information about the physical state of
matter must be lost as the matter is sucked into a BH, otherwise, this information would
distinguish one BH from another. In 1974, Hawking made the landmark conjecture that BHs
do not simply suck in everything, but rather behave as black bodies that emit radiation
as well as absorbing it \cite{Hawking:1974sw}. He calculated the black-body temperature of a BH known as Hawking
temperature. Having a distinct temperature implies that a BH has entropy, which Hawking also
calculated. Entropy is a measure of the number of different ways the microscopic constituents
of a BH can arrange themselves. This goes against the no-hair theorem, which says that a BH
can only be arranged in one way  as defined by its mass and spin. Recently, Hawking with
his colleagues suggested that information-preserving massless particles known as soft hair
could surround BHs. They have calculated the entropy of a BH that has a certain kind of soft
hair, which leads to the Hawkings original calculation of BH entropy \cite{H1}.

Einstein-Maxwell-dilaton (EMD) theory is an arousing and highly motivated theory
to examine the affect of new essential degrees of freedom in order to prove the
no-hair theorem because of the existence of hairy BH solutions in contact 
with the vector, scalar and tensor spectrums. It is to be observed that, in the background of asymptotically
anti de Sitter (AdS) geometries, there must exist exact
scalar-hairy BH solutions with respect to specific scalar potentials \cite{Sudarsky:2002mk,Nucamendi:1995ex}.

At the back of no-hair conjecture, the simplest physical visualization is: after the 
gravitational collapse, the matter fields left over in the exterior region would finally 
be immersed by the BH itself or be radiated off to infinity. Although, the hair of a BH 
is located at the outer side of event horizon, the question then arises that 
how there exist a possibility that the matter can hover in an immense gravitational field
without collapsing entirely? Intuitively, the respond to the question could be that this is possible
if the internal pressure is large up to a sufficient degree. On the other hand, such type of intuition does work
only close to the event horizon and it must be noted that the non-linear interaction of matter fields 
provides the basis of the existence of hairy BHs \cite{Nunez:1996xv}.

Geometric spontaneous scalarisation phenomenon has recently been examined \cite{Doneva:2017bvd,Silva:2017uqg,Cunha:2019dwb,Blazquez-Salcedo:2018jnn,Herdeiro:2018wub}. 
It is noted that in gravitational field, matter models where a real/scalar field 
couples minimally to the curvature squared Gauss-Bonnet (GB) combination under 
particular options of the coupling function, there exists two possibilities of solutions. The standard vacuum BH solutions (bald) of general relativity (GR)
and very recent "hairy" BH solutions with a scalar field characterization, both types of solutions deceiving 
no-scalar hair conjecture. Moreover, it is suggested that the hairy BHs could form
by means of spontaneous scalarisation, since the standard vacuum BH solutions 
were shown to be perturbatively unstable. Following the previous analysis, Herdeiro and Radu \cite{Herdeiro:2015waa,Herdeiro:2019yjy} have examined the basics of the spontaneous
scalarisation phenomenon for BHs. They have elaborated this phenomenon within the
quantum framework by giving two examples of spherically symmetric
static asymptotically flat BHs in effective field theory. One of the example is
that the trace anomaly occurs in the matter sector: provides a generalized form of the 
Reissner-Nordstr\"{o}m BH solution incorporating $F^{4}$ correction. While,
in second case it arises from the geometry sector: provides geometrically a 
non-commutative generalized Schwarzschild BH. In comparison, they have 
also discussed the scalarisation phenomenon of Einstein-Maxwell-dilaton BHs. Because of
a non-minimal coupling, they have investigated the connection of the scalarisation mechanism with the quantum instabilities \cite{Mendes:2013ija}.

An important fundamental predication of GR is the phenomenon of gravitational lensing, which is the deflection
of light in the presence of gravitational fields of compact objects.
Solender $(1801)$ proposed the gravitational lensing due to the Sun in the background 
of Newton's theory, while the absolute value of deflection angle $\delta=4M_{\odot}/R_{\odot}\simeq1.75$
arc-sec was obtained in the background of GR by Chowlson $(1929)$ and Einstein $(1936)$.
In depth, the study of gravitational lensing of compact objects initiated with the learning of facts
of the double quasar $Q0957+561$ having redshift $\mathcal{Z}=0.39$ (Young et al. $1980$) \cite{young,VallsGabaud:2012xz,Bartelmann:1999yn}. They focused on
the effects of the weak gravitational lensing, i.e., the lensing for which the deflection angle is very small,
equal to a few arc-seconds. Generally, there exist a differentiation between lensing on non-cosmological
as well as on cosmological distances. Moreover, the assumption of luminosity
distances which are model-dependent leads us to the opportunity to examine
the mass distribution of dark matter haloes, in particular the inner density slopes \cite{Cunha:2015yba,Bozza:2009yw,Bartelmann:2010fz}. Schunck et al. \cite{Schunck:2006rk} have investigated the deflection angle of
a spherically symmetric halo built from a precisely solvable scalar field
model incorporating Emden type self-interaction. They have examined the gravitational lensing impact
by considering bosonic configuration and obtained
the normalized projected mass as well as the corrections with respect to the pressure
and analyze the weak-field dimension-less lens equation.

In this paper, we try to understand the effect of the hair on deflection angle by asymptotically flat black holes in Einstein-Maxwell-dilaton (EMD) theory, where is  derived from a string theory at low energy limits \cite{C1}. The EMD black holes have a scalar hair, in addition to mass, rotation and charge so that they offers an fascinating theoretical black hole models to check experimentally in black hole experiments, and investigate possible differences between Einstein gravity and modified gravity theories \cite{C1,B17,A4}.

Further, let us now briefly review the GBT which connects the topologically surfaces. First, using Euler characteristic $\chi$ and a Riemannian metric $g$, one can choose the subset oriented surface domain as $(D,\chi,g)$ to find the Gaussian curvature $K$. Then the Gauss-Bonnet theorem is defined as follow  \cite{A6}
\begin{equation}
\iint_{D} K \mathrm{d} S+\oint_{\partial D} \kappa \mathrm{d} t+\sum_{i} \theta_{i}=2 \pi \chi\left(D\right),
\label{gb}
\end{equation} where $\kappa$ is the geodesic curvature for  $\partial D:\{t\}\rightarrow D$ and $\theta_i$ is the exterior angle with $i^{th}$ vertex. Following this approach, global symmetric lenses are considered to be Riemannian metric manifolds, which are geodesic spatial light rays. In optical geometry, we calculate the Gaussian optical curvature $K$ to find the asymptotic bending angle which can be calculated as follows \cite{A6}: 
\begin{equation}
\hat{\alpha}=-\int \int_{D_\infty} K  \mathrm{d}S.
\end{equation}

Note that this equation is an exact result for the deflection angle. In this equation, we integrate over an infinite region of the surface $D_\infty$ which is bounded by the light ray. By assumption, one can use the above relation only for asymptotically Euclidean optical metrics. Therefore it will be interesting to see the form of the deflection angle in the case of non-asymptotically Euclidean metrics. This method has been applied in various papers for different types of spacetimes \cite{A7,Crisnejo:2019xtp,R5,Crisnejo:2018ppm,R6,Arakida:2017hrm,Ono:2018ybw,Jusufi:2017vta,Ovgun:2018xys,Jusufi:2017mav,Ono:2017pie,Jusufi:2017vew,Ono:2018jrv,Jusufi:2017lsl,Ovgun:2019wej,Jusufi:2017hed,Sakalli:2017ewb,Ovgun:2018prw,Jusufi:2017uhh,Kumaran:2019qqp,Jusufi:2018jof,Ovgun:2018fnk,Ovgun:2018ran,Ovgun:2018oxk,Ovgun:2018fte,Ovgun:2018tua,Javed:2019qyg,Javed:2019a,Javed:2019b}.

This paper is organized as follows: In section \textbf{$\textrm{2}$}, we review some
basic concepts about asymptotically flat hairy BH. In section \textbf{$\textrm{3}$}, we compute the Gaussian optical curvature for
deflection angle and calculate the deflection angle by using GBT for $\gamma=1$ also find the deflection angle in a plasma medium.
In section \textbf{$\textrm{4}$}, we calculate the deflection angle for $\gamma=\sqrt{3}$ also find the deflection angle in a plasma medium. The last section comprises of concluding remarks and results obtained from graphical analysis.

\section{Asymptotically flat black holes in Einstein-Maxwell-dilaton theory}
In this section, we briefly review the asymptotically flat hairy BH in EMD theory in the
background of dilaton potential. We consider the following action
of Einstein-Maxwell-dilaton theory \cite{C1,B17}:
\begin{equation}
I[g_{\mu\nu},A_{\mu},\phi]=\frac{1}{2k}\int_{\mathcal{M}}d^{4}x\sqrt{-g}[R-e^{\gamma\varphi}F^{2}-
\frac{1}{2}(\partial\varphi)^{2}-V(\varphi)],\label{j1}
\end{equation}
where
\begin{equation}
F^{2}=F_{\mu\nu}F^{\mu\nu}, (\partial\varphi)^{2}=\partial_{\mu}\varphi\partial^{\mu}\varphi,
\end{equation}
$V(\varphi)$ is the dilaton potential, and $c=G_{N}=4\pi\epsilon_{0}=1$ by using the convention $k=8\pi$.
The equations of motion for gauge field, dilaton and metric are following:
\begin{eqnarray}
R_{\mu\nu}-\frac{1}{2}g_{\mu\nu}R&=&T_{\mu\nu}^{\varphi}+T^{EM},\\
\partial_{\mu}(\sqrt{-g}e^{\gamma\varphi}F^{\mu\nu})&=&0,\\
\frac{1}{\sqrt{-g}}\partial_{\mu}(\sqrt{-g}g^{\mu\nu}\partial_{\nu}\varphi)&=&\frac{dV(\varphi)}{d\varphi}+
\gamma e^{\gamma\phi}F^{2},
\end{eqnarray}
where the stress tensors of matter fields are defined as \begin{equation}T_{\mu\nu}^{\varphi}\equiv\frac{1}{2}
\partial_{\mu}\phi\partial_{\nu}\varphi-\frac{1}{2}g_{\mu\nu}[\frac{1}{2}(\partial\phi)^{2}+V(\phi)],\end{equation}
\begin{equation}T^{EM}_{\mu\nu}\equiv 2e^{\gamma\varphi}(F_{\mu\alpha}F^{\alpha}_{\nu}-\frac{1}{4}g_{\mu\nu}F^{2}).\end{equation}

For exact regular hairy BH solutions, for a general scalar potential there is a new method
in a number of papers \cite{A4}, by using a specific ansatz. For
flat spacetime, we should apply the similar special ansatz for
the metric and dilaton. For the sake of simplicity, we are going to focus on two particular cases
for which the exponent coefficient of the dilaton coupling with the gauge field takes the values
$\gamma=1$ and $\gamma=\sqrt{3}$, but for more general solution there exist a literature \cite{B17}.

\section{Weak Deflection Angle of Calculation of photon lensing for $\gamma=1$ by Gauss-Bonnet Theorem}
For this solution we consider the following scalar field potential:
\begin{equation}
V(\varphi)=2\alpha(2\varphi+\varphi\cosh\varphi-3\sinh\varphi),
\end{equation}
here $\alpha$ is an arbitrary parameter.
For $\gamma=1$ the static hairy BH metric and gauge field, yields that:
\begin{eqnarray}
ds^{2}&=&\Omega(x)\left[-f(x)dt^{2}+\frac{\eta^{2}dx^{2}}{x^{2}f(x)}+d\Sigma^{2}\right],\\
F&=&\frac{1}{2}F_{\mu\nu}dx^{\mu}\wedge dx^{\nu}=\frac{qe^{-\phi}}{x}dt\wedge dx,\nonumber
\end{eqnarray}
where $\eta$ and $q$ are defined as independent parameters of the solution and correspond to the
mass and charge of this BH. $d\Sigma^{2}=d\theta^{2}+\sin^{2}\theta d\phi^{2}$ is spherical line element,
and the coordinate $x$ is restricted to be positive, $x\in[0,\infty)$. We can suppose that $\eta>0$. One can use the conformal factor as follows:
\begin{equation}
\Omega(x)=\frac{x}{\eta^{2}(x-1)^{2}},
\end{equation}
and then check that the equations of motion are satisfied for the following spacetime metric function:
\begin{equation}
f(x)=\alpha\left[\frac{x^{2}-1}{2x}-\ln(x)\right]+\frac{\eta^{2}(x-1)^{2}}{x}\left[1-\frac{2q^{2}(x-1)}{x}\right].
\end{equation}

It is appropriate to first find the black hole optical metric by imposing
the null condition $ds^{2}=0$, and solving the space-time metric for $dt$
and also set the metric into equatorial plane with $\theta=\frac{\pi}{2}$,
which yield as:
\begin{equation}
dt^{2}=\frac{\eta^{2}}{x^{2}f(x)^{2} }dx^{2}+\frac{1}{f(x)}d\varphi^{2}.
\end{equation}
The optical geometry is in two dimensions, and is obtained for thermodynamically
stable asymptotically flat hairy BH with a dilaton potential as follows \cite{A6}.
By using Gauss-Bonnet theorem, initially we find the Gaussian curvature $\mathcal{K}$
of the optical spacetime, as
\begin{equation}
\mathcal{K}=\frac{R_{icciScalar}}{2},\label{j2}
\end{equation}

\begin{eqnarray}
\mathcal{K}\approx -{\eta}^{2}x-{\frac {{\eta}^{2}}{x}}+1/4\,{\frac {{\eta}^{2}}{{x}^{2}}
}+3/2\,{\eta}^{2}+1/4\,{x}^{2}{\eta}^{2}-1/2\,x\alpha\,\ln  \left( x
 \right) -1/2\,{\frac {\alpha\,\ln  \left( x \right) }{x}}+1/4\,\alpha
\,{x}^{2}-1/4\,{\frac {\alpha}{{x}^{2}}}-{x}^{2}{\eta}^{2}{q}^{2} \notag\\+5\,{
\eta}^{2}{q}^{2}x-11\,{\frac {{\eta}^{2}{q}^{2}}{{x}^{2}}}+16\,{\frac 
{{\eta}^{2}{q}^{2}}{x}}+3\,{\frac {{\eta}^{2}{q}^{2}}{{x}^{3}}}-12\,{
\eta}^{2}{q}^{2}+5\,{\frac {\alpha\,{q}^{2}}{x}}-3/2\,{\frac {\alpha\,
{q}^{2}}{{x}^{3}}}\notag\\-4\,{\frac {\alpha\,{q}^{2}\ln  \left( x \right) }{{
x}^{2}}}+3\,{\frac {\alpha\,{q}^{2}\ln  \left( x\right) }{x}}+x\alpha
\,\ln  \left( x \right) {q}^{2}-3\,\alpha\,{q}^{2}-1/2\,{x}^{2}\alpha
\,{q}^{2}+1/2\,x\alpha\,{q}^{2}-1/2\,{\frac {\alpha\,{q}^{2}}{{x}^{2}}
}.
\end{eqnarray}
For multiple images, we use the global theory (Gauss-Bonnet theorem) to relate with the local
feature of the space-time such that Gaussian optical curvature.

The above equation will be apply to calculate the deflection angle by taking a non singular domain
$S_{R}$ outside of the light ray (along with boundaries $\partial S_{R}=\gamma_{g}\cup C_{R}$)
with Euler characteristic $\chi(S_{R})$, Gaussian curvature $\mathcal{K}$, geodesic curvature $k$
and exterior jump angles $\alpha_{i}=(\alpha_{O}, \alpha_{S})$ at vertices.
\begin{equation}
\int\int_{S_{R}}\mathcal{K}dS+\oint_{\partial S_{R}}k dt+\sum_{j}\alpha_{j}=2\pi\chi(S_{R}),
\end{equation}
at weak limit approximation
$(\rho\rightarrow\infty)$, $\alpha_{O}+\alpha_{S}\rightarrow\pi$. Then GBT becomes
\begin{equation}
\int\int_{S_{R}}\mathcal{K}dS+\oint_{C_{r}}kdt =^{\rho\rightarrow\infty}\int\int_{S_{\infty}}kdS+\int^{\pi+\Theta}_{0}d\varphi=\pi.
\end{equation}
Now, by geodesic property, the geodesic curvature vanishes $k(\gamma_{g})=0$, and we get
\begin{equation}
k(C_{R})=|\nabla_{\dot{C}_{r}\dot{C}_{r}}|,
\end{equation}
with $C_{r}:=\rho(\varphi)=r=$constant. Then GBT reduces
\begin{equation}
\lim_{R\rightarrow\infty}\int^{\pi+\Theta}_{0}\left[k_{g}\frac{d\sigma}{d\varphi}\right]|_{C_{R}}d\varphi=\pi-\lim_{R\rightarrow\infty}
\int\int_{S_{R}}\mathcal{K}dS.
\end{equation}
Now, for radial distance
\begin{equation}
k(C_{R})dt=d\varphi.
\end{equation}
Therefore,
\begin{equation}
\lim_{R\rightarrow\infty}k_{g}\frac{d\sigma}{d\varphi}|_{C_{R}}=1.
\end{equation}
In the weak field regions, the light ray follows a straight line approximation,
so that we can use the condition of $r=b/\sin\varphi$
at zero order.
\begin{equation}
\Theta=-\lim_{R\rightarrow\infty}\int^{\pi}_{0}\int^{R}_{b/\sin\varphi}\mathcal{K}dS,
\end{equation}

where \begin{equation}K dS \approx  1/4\,{\frac {{\eta}^{2}}{{x}^{2}}}+1/4\,{\frac {\alpha}{{x}^{2}}}-{
\frac {{\eta}^{2}{q}^{2}}{{x}^{2}}}-1/2\,{\frac {\alpha\,{q}^{2}}{{x}^
{2}}}.
\end{equation}
After simplification, we find the deflection angle of photon for asymptotically flat hairy BH in leading order terms as
\begin{equation}
\tilde{\alpha} \simeq -1/2\,{\frac {{\eta}^{2}}{b}}-1/2\,{\frac {\alpha}{b}}+2\,{\frac {{
\eta}^{2}{q}^{2}}{b}}+{\frac {\alpha\,{q}^{2}}{b}}. \label{sol0}
\end{equation}
Therefore, we can say that GBT provides a globally as well as topologically effect, this method is very useful 
for quantitative tool and can be apply in any asymptotically flat metrics.
\subsection{Photon lensing in a plasma medium}
In this section, we analyze the effect of plasma medium on the photon lensing by asymptotically hairy black hole.
The refractive index for hairy black hole is as follows \cite{A7},
\begin{equation}
n(x)=\sqrt{1-\frac{\omega_{e}^{2}}{\omega_{\infty}^{2}}\left(\frac{xf(x)}{\eta^{2}(x-1)^{2}}\right)},
\end{equation}
then, the corresponding optical metric yields that
\begin{equation}
d\tilde{\sigma}^{2}=g^{opt}_{jk}dx^{j}dx^{k}=\frac{n^{2}(x)}{f(x)}\left(\frac{\eta^{2}}{x^{2}f(x)}dx^{2}+d\varphi^{2}\right).
\end{equation}
The determinant of above optical metric is:
\begin{equation}
detg^{opt}_{x\varphi}=\frac{xf(x)\omega_{e}^{4}-2\eta^{2}\omega_{\infty}^{2}\omega_{e}^{2}(x-1)^{2}}{\eta^{2}\omega_{\infty}^{4}xf(x)^{2}(x-1)^{4}},
\end{equation} 
where the metric function is given as
\begin{equation}
f(x)=\alpha\left[\frac{x^{2}-1}{2x}-\ln(x)\right]+\frac{\eta^{2}(x-1)^{2}}{x}\left[1-\frac{2q^{2}(x-1)}{x}\right].
\end{equation}

Now, we have
\begin{equation}
\frac{d\tilde{\sigma}}{d\varphi}=n(x)\left(\frac{\alpha^{2}x^{2}}{f(x)}\right)^{1/2},
\end{equation}
hence we get differently which goes to $\alpha$:
\begin{equation}
\lim_{x\rightarrow\infty}k_{g}\frac{d\tilde{\sigma}}{d\varphi}|_{C_{R}}=1.
\end{equation}
We use straight line approximation $r=b/\sin\varphi$, for the limit $x\rightarrow\infty$, then GBT stated as
\begin{equation}
\lim_{x\rightarrow\infty}\int^{\pi+\Theta}_{0}\left[k_{g}\frac{d\tilde{\sigma}}{d\varphi}\right]|_{C_{R}}d\varphi=\pi-
\lim_{x\rightarrow\infty}\int^{\pi}_{0}\int^{x}_{b/\sin\varphi}\mathcal{K}dS,
\end{equation}
where 
\begin{eqnarray}
\mathcal{K}dS&=&-3/2\,{\frac {{\omega_{e}}^{2}{\eta}^{2}{q}^{2}}{{x}^{2}{\omega_{\infty}}^{2}}}-3/2\,{\frac {{\omega_{e}}^{2}\alpha\,{q}^{2}}{{
x}^{2}{\omega_{\infty}}^{2}}}+1/4\,{\frac {{\omega_{e}}^{2}{
\eta}^{2}}{{x}^{2}{\omega_{\infty}}^{2}}}+3/8\,{\frac {{\omega_{\infty}}^{2}\alpha}{{x}^{2}{\omega_{\infty}}^{2}}}+1/4\,{\frac {\alpha}{
{x}^{2}}}+1/4\,{\frac {{\eta}^{2}}{{x}^{2}}}-1/2\,{\frac {\alpha\,{q}^
{2}}{{x}^{2}}}-{\frac {{\eta}^{2}{q}^{2}}{{x}^{2}}}.
\end{eqnarray}
Note that we only consider the first order terms. 
After simplification, we obtain the deflection angle in weak field limits as follows:
\begin{equation}
\tilde{\alpha}\simeq 
3\,{\frac {{\omega_{e}}^{2}{\eta}^{2}{q}^{2}}{b{\omega_{\infty}
}^{2}}}+2\,{\frac {{\eta}^{2}{q}^{2}}{b}}+3\,{\frac {{\omega_{e}
}^{2}\alpha\,{q}^{2}}{b{\omega_{\infty}}^{2}}}+{\frac {\alpha\,{q}^{2
}}{b}}-1/2\,{\frac {{\omega_{e}}^{2}{\eta}^{2}}{b{\omega_{\infty}}^{2}}}-1/2\,{\frac {{\eta}^{2}}{b}}-3/4\,{\frac {{\omega_{e}}^{2}\alpha}{b{\omega_{\infty}}^{2}}}-1/2\,{\frac {\alpha}{b}}. \label{sol00}
\end{equation}
The above results shows that the photon rays are moving in a medium of homogeneous plasma.   

\section{New Asymptotically flat black holes in Einstein-Maxwell-dilaton theory}
In this section, we analyzed the exact asymptotically flat charged hairy BH's in the
background of dilaton potential. There exist a literature related to this BH recently
discussed by Astefanesei et al. \cite{B17}. We are interested in a action
of of Einstein-Maxwell-dilaton theory ($\kappa=8\pi G_{N}$):
\begin{equation}
I[g_{\mu\nu},A_{\mu},\phi]=\frac{1}{2\kappa}\int d^{4}x\sqrt{-g}\left[
R-\frac{1}{4}e^{\gamma\phi}F^{2}-\frac{1}{2}\partial_{\mu}\phi\partial^{\mu
}\phi-V(\phi)\right],
\end{equation}
where the gauge coupling and potential are functions of the dilaton. The corresponding equations of motion are
\begin{equation}
\nabla_{\mu}\left(  e^{\gamma\phi}F^{\mu\nu}\right)  =0,
\end{equation}%
\begin{equation}
\frac{1}{\sqrt{-g}}\partial_{\mu}\left(  \sqrt{-g}g^{\mu\nu}\partial_{\nu}%
\phi\right)  -\frac{\partial V}{\partial\phi}-\frac{1}{4}\gamma e^{\gamma\phi
}F^{2}=0,\label{dil}%
\end{equation}%
\begin{equation}
R_{\mu\nu}-\frac{1}{2}g_{\mu\nu}R=\frac{1}{2}\left[  T_{\mu\nu}^{\phi}%
+T_{\mu\nu}^{EM}\right],
\end{equation}
where $T_{\mu\nu}^{\phi}$ and $T_{\mu\nu}^{EM}$ are the stress tensors of the matter fields
and represented as follows
\begin{equation}
T_{\mu\nu}^{\phi}=\partial_{\mu}\phi\partial_{\nu}\phi-g_{\mu\nu}\left[
\frac{1}{2}\left(  \partial\phi\right)  ^{2}+V(\phi)\right]
\,\,\,\,,\,\,\,\,\,T_{\mu\nu}^{EM}=e^{\gamma\phi}\left(  F_{\mu\alpha}F_{\nu
}^{\cdot\alpha}-\frac{1}{4}g_{\mu\nu}F^{2}\right).
\end{equation}

Now, we consider only asymptotically flat solutions. To implement
this condition, we require
\begin{equation}
\lim_{x\rightarrow1}\Omega(x)f(x)=1, \label{BC1}%
\end{equation}
on this account, we fix $f_{0}$ to obtain \cite{B17}
\begin{align}
f(x)  &  =\frac{\eta^{2}}{\nu}\left(  x^{2}+\frac{2x^{2-\nu}}{\nu-2}-\frac
{\nu}{\nu-2}\right)  +f_{1}\left(  \frac{x^{\nu+2}}{\nu+2}-x^{2}%
+\frac{x^{2-\nu}}{2-\nu}+\frac{\nu^{2}}{\nu^{2}-4}\right) \nonumber\\
&  {+}\frac{Q^{2}\eta^{2}}{(1-p)\nu^{2}}\left(  \frac{x^{3-p+\nu}}{3-p+\nu
}+\frac{x^{3-p-\nu}}{3-p-\nu}-2\frac{x^{3-p}}{3-p}-\frac{2\nu^{2}}{\left(
3-p\right)  \left(  3-p+\nu\right)  \left(  3-p-\nu\right)  }\right)
\nonumber\\
&  {+}\frac{P^{2}\eta^{4}}{(1+p)\nu^{2}}\left(  \frac{x^{3+p+\nu}}{3+p+\nu
}+\frac{x^{3+p-\nu}}{3+p-\nu}-2\frac{x^{3+p}}{3+p}-\frac{2\nu^{2}}{\left(
3+p\right)  \left(  3+p+\nu\right)  \left(  3+p-\nu\right)  }\right).
\label{general}%
\end{align}

\subsection{Solutions with a non-trivial dilaton potential ($\gamma=1$)}

For this solution we consider the following scalar field potential \cite{B17}:
\begin{equation}
\Omega(x)=\frac{x}{\eta^{2}\left(  x-1\right)  ^{2}} \,\,\,\,\,\,\,\,\, ,
\,\,\,\,\,\, \qquad\phi(x)=\ln(x), \label{dilaton}%
\end{equation}%
\begin{equation}
ds^{2}=\Omega(x)\left[  -f(x)dt^{2}+\frac{\eta^{2}dx^{2}}{x^{2}f(x)}%
+d\theta^{2}+\sin^{2}{\theta}d\varphi^{2}\right].
\end{equation}

Here we study only the $\gamma=1$ case, which is smoothly connected with a solution
of $\mathcal{N}=4$ supergravity which gives the solutions of metric function as follows:
\begin{equation}
A=\frac{Q}{x}dt+P\cos\theta d\varphi, \label{A}%
\end{equation}%
\begin{equation}
V(\phi)=\alpha\left[  2\phi+\phi\cosh(\phi)-3\sinh(\phi)\right],
\label{potential1}%
\end{equation}%
\begin{equation}
f(x)=\frac{\eta^{2}(x-1)^{2}}{x}+\left[  \frac{x}{4}-\frac{1}{4x}-\frac{1}%
{2}\ln(x)\right]  \alpha+\frac{\eta^{2}(x-1)^{3}}{2x}\left(  \eta^{2}%
P^{2}-x^{-1}Q^{2}\right).
\end{equation}
Note that the dilaton field is vanishing at the boundary, $x=1$ and also the dilaton potential is
vanishing when $\alpha=0$. At the boundary $x=1$, one can find  a asymptotically flat spacetime. After make a change of coordinates using:
\begin{equation}
\Omega(x)=r^{2}+O(r^{-4}), \label{AF}%
\end{equation}
which is given for, the $x<1$ black holes, by
\begin{equation}
x=1-\frac{1}{\eta r}+\frac{1}{2\eta^{2}r^{2}}-\frac{1}{8\eta^{3}r^{3}}%
+\frac{1}{2^{7}\eta^{5}r^{5}}. \label{CC}%
\end{equation}

The spacetime metric becomes asymptotically flat \cite{B17}%

\begin{equation}
g_{tt}=\Omega(x)f(x)=1-\frac{\alpha+6\eta^{2}\left(  \eta^{2}P^{2}%
-Q^{2}\right)  }{12\eta^{3}r}+O(r^{-2}),
\end{equation}

\begin{equation}
g_{rr}^{-1}=\frac{x^{2}f(x)}{\eta^{2}\Omega(x)}\left(  \frac{dr}{dx}\right)
^{2}=1-\frac{\alpha+6\eta^{2}\left(  \eta^{2}P^{2}-Q^{2}\right)  }{12\eta
^{3}r}+O(r^{-2}).
\end{equation}

It is noted that the scalar field potential is regular everywhere, except at the spacetime singularities.

\section{Deflection angle of photons by asymptotically flat hairy black holes in Einstein-Maxwell-dilaton theory}

 Initially we find the Gaussian curvature $\mathcal{K}$
of the optical spacetime, as
\begin{equation}
\mathcal{K}=\frac{R_{icciScalar}}{2},
\end{equation}
and
\begin{equation}
    \mathcal{K}={\frac { \left( -6\,{\eta}^{4}{p}^{2}+6\,{Q}^{2}{\eta}^{2}-\alpha
 \right)  \left( -6\,{\eta}^{4}{p}^{2}+6\,{Q}^{2}{\eta}^{2}+16\,{\eta}
^{3}r-\alpha \right) }{192\,{\eta}^{6}{r}^{4}}}.
\end{equation}

In weak field limits,
\begin{equation}
    \mathcal{K}=-\,{\frac {\alpha}{{12\eta}^{3}{r}^{3}}}+{\frac {{\alpha}^{2}}{192\,
{\eta}^{6}{r}^{4}}}+\,{\frac { \left( -8\,{\eta}^{3}r+\alpha
 \right) {p}^{2}}{{16\eta}^{2}{r}^{4}}}-\,{\frac { \left( -8\,{\eta}
^{3}r+\alpha \right) {Q}^{2}}{{16\eta}^{4}{r}^{4}}}+\mathcal{O}
( {Q}^{3},p^3 ).
\end{equation}
For multiple images, we use the global theory (Gauss-Bonnet theorem)
to relate with the local feature of the space-time such that Gaussian
optical curvature.

In the weak field regions, the light ray follows a straight line approximation,
so that we can use the condition of $r=b/\sin\phi$
at zero order.
\begin{equation}
\tilde{\alpha}=-\lim_{R\rightarrow\infty}\int^{\pi}_{0}\int^{R}_{b/\sin\varphi}\mathcal{K}dS.
\end{equation}
Now by using Eq. $(18)$, the deflection angle of photon by exact asymptotically flat
charged hairy black hole with a dilaton potential in weak field limit is found as:
\begin{eqnarray}
    \tilde{\alpha}&=&{\frac {3\,{Q}^{2}{p}^{2}\pi}{32\,{b}^{2}}}+{\frac {\eta\,{p}^{2}}{b}}-{\frac {{Q}^{2}}{b\eta}}+{\frac {\pi\,{Q}^{2}\alpha}{64\,{b}^{
2}{\eta}^{4}}}+\,{\frac {\alpha}{6b{\eta}^{3}}}+\mathcal{O}
( {Q}^{3},p^3 ). \label{sol1}
\end{eqnarray}

\subsection{Deflection angle of photons in plasma medium by asymptotically flat hairy black holes in Einstein-Maxwell-dilaton theory}
In this section, we analyze the effect of plasma medium on the photon
lensing by asymptotically hairy black hole. The refractive index for
hairy black hole is as follows \cite{A7},
\begin{equation}
n(x)=\sqrt{1-\frac{{\omega_{e}}^{2}}{{\omega_{\infty}}^{2}}\left(\frac{xf(x)}
{\eta^{2}(x-1)^{2}}\right)},
\end{equation}
then, the corresponding optical metric yields that
\begin{equation}
d\tilde{\sigma}^{2}=g^{opt}_{jk}dx^{j}dx^{k}=\frac{n^{2}(x)}{f(x)}
\left(\frac{\eta^{2}}{x^{2}f(x)}dx^{2}+d\varphi^{2}\right).
\end{equation}
The determinant of above optical metric is:
\begin{equation}
detg^{opt}_{x\varphi}=\frac{xf(x){\omega_{e}}^{4}-2\eta^{2}{\omega_{\infty}}^{2}
{\omega_{e}}^{2}(x-1)^{2}}{\eta^{2}{\omega_{\infty}}^{4}xf(x)^{2}(x-1)^{4}}.
\end{equation}

Now, by using Eq. $17$ and $29$, in weak field limit the Gaussian
optical curvature stated as follows:
\begin{eqnarray}
\mathcal{K}&=&-\,{\frac {{\omega_{\infty}}^{2} \left( {\omega_{e}}^{2}-2\,
{\omega_{\infty}}^{2} \right) {Q}^{2}}{4{r}^{3}\eta\, \left( {\omega_{e}}^{2}-{\omega_{\infty}}^{2} \right) ^{2}}}+\,{\frac {{
\omega_{\infty}}^{2} \left( 3\,{Q}^{2}{\omega_{e}}^{2}{\omega_{\infty}}^{2}+3\,{Q}^{2}{\omega_{\infty}}^{4}+2\,\eta\,r{\omega_{e}}^{4}-6\,\eta\,r{\omega_{e}}^{2}{\omega_{\infty}}^{2}+4\,
\eta\,r{\omega_{\infty}}^{4} \right) {p}^{2}}{8{r}^{4} \left( {\omega_{e}}^{2}-{\omega_{\infty}}^{2} \right) ^{3}}}\nonumber\\&+&\,{\frac {
\alpha\, \left( \Psi \right) {\omega_{\infty}}^{2}}{48{\eta}^{4}{r}^{5} \left( {\omega_{e}}^{2}-{\omega_{\infty}}^{2} \right) ^{4}}},
\end{eqnarray}
where \begin{eqnarray}
    \Psi=9\,{Q}^{2}\eta\,{p}^{2}{\omega_{e}}^{4}{\omega_{\infty}}^{2}+9\,{Q}^{2}\eta\,{p}^{2}{\omega_{e}}^{2}{\omega_{\infty}}^{4}-3\,{\eta}^{2}{p}^{2}r{\omega_{e}}^{4}{\omega_{\infty}}^{2}+3\,{\eta}^{2}{p}^{2}r{\omega_{\infty}}^{6}+3\,{Q}^{2}r{
\omega_{e}}^{4}{\omega_{\infty}}^{2}-3\,{Q}^{2}r{\omega_{\infty}}^{6}+2\,\eta\,{r}^{2}{\omega_{e}}^{6} \notag\\-8\,\eta\,{r}^{2}{
\omega_{e}}^{4}{\omega_{\infty}}^{2}+10\,\eta\,{r}^{2}{\omega_{e}}^{2}{\omega_{\infty}}^{4}-4\,\eta\,{r}^{2}{\omega_{\infty}}
^{6}.
\end{eqnarray}
Now, we have
\begin{equation}
\frac{d\tilde{\sigma}}{d\varphi}=n(x)\left(\frac{\alpha^{2}x^{2}}{f(x)}\right)^{1/2},
\end{equation}
hence we get differently which goes to $\alpha$:
\begin{equation}
\lim_{x\rightarrow\infty}k_{g}\frac{d\tilde{\sigma}}{d\varphi}|_{C_{R}}=\alpha.
\end{equation}
We use straight line approximation $r=b/\sin\varphi$, for the limit $x\rightarrow\infty$, then GBT stated as
\begin{equation}
\lim_{x\rightarrow\infty}\int^{\pi+\tilde{\alpha}}_{0}\left[k_{g}\frac{d\tilde{\sigma}}
{d\varphi}\right]|_{C_{R}}d\varphi=\pi-\lim_{x\rightarrow\infty}\int^{\pi}_{0}
\int^{x}_{b/\sin\varphi}\mathcal{K}dS.
\end{equation}
After simplification, we obtain
\begin{eqnarray}
    \tilde{\alpha}&=&{\frac {21\,\pi\,{Q}^{2}{p}^{2}{\omega_{e}}^{2}}{64\,{\omega_{\infty}}^{2}{b}^{2}}}+{\frac {3\,{Q}^{2}{p}^{2}\pi}{32\,{b}^{2}}}+{
\frac {\eta\,{p}^{2}{\omega_{e}}^{2}}{{\omega_{\infty}}^{2}b}}+
{\frac {\eta\,{p}^{2}}{b}}-{\frac {7\,\pi\,\alpha\,{p}^{2}{\omega_{e}}^{2}}{128\,{\omega_{\infty}}^{2}{b}^{2}{\eta}^{2}}}-{\frac {
\pi\,\alpha\,{p}^{2}}{64\,{b}^{2}{\eta}^{2}}}-{\frac {{Q}^{2}{\omega_{e}}^{2}}{{\omega_{\infty}}^{2}b\eta}}\nonumber\\&-&{\frac {{Q}^{2}}{b\eta}}+
{\frac {7\,\pi\,{Q}^{2}\alpha\,{\omega_{e}}^{2}}{128\,{\omega_{\infty}}^{2}{\eta}^{4}{b}^{2}}}+{\frac {\pi\,{Q}^{2}\alpha}{64\,{b}^{
2}{\eta}^{4}}}+1/6\,{\frac {\alpha\,{\omega_{e}}^{2}}{{\omega_{\infty}}^{2}{\eta}^{3}b}}+1/6\,{\frac {\alpha}{b{\eta}^{3}}}. \label{sol2}
\end{eqnarray}
The proposed deflection angle shows that the photon rays are
moving in a medium of homogeneous plasma.

\section{Conclusion}
In this paper, we obtain the deflection angle of photon to the
spherically symmetric and asymptotically flat spacetime of hairy
BH with Einstein-Maxwell-dilaton system in weak field limit. To
this end, we set the photon rays on the equatorial plane in the
black hole spacetime. For this purpose, we have used the GBT
and obtain the deflection angle of photon by integrating a domain
outside the impact parameter. Moreover, we also found the deflection
angle of photon by asymptotically flat hairy BH in plasma medium. We
examined that the proposed deflection angle shows that gravitational
lensing can be affected from the hair of the black hole and it is a global effect as well as a valuable tool to study the nature
of singularities of black holes.

In our analysis, we obtain the deflection angle of photon to the spherically symmetric and
asymptotically flat spacetime of hairy BH with Einstein-Maxwell-dilaton (EMD) system in the weak-field limit by using
Gauss-Bonnet theorem.
The computed deflection angle defined by \eqref{sol1} is stated as follows
\begin{eqnarray}
\tilde{\alpha}&=&{\frac {3\,{Q}^{2}{p}^{2}\pi}{32\,{b}^{2}}}+{\frac {\eta\,{p}^{2}}{b}}-
{\frac{{Q}^{2}}{b\eta}}+{\frac {\pi\,{Q}^{2}\alpha}{64\,{b}^{2}{\eta}^{4}}}+\,
{\frac {\alpha}{6b{\eta}^{3}}}+\mathcal{O}({Q}^{3},p^3).\label{WW1}
\end{eqnarray}

It is noted that, for the selection of mass term $\eta=m$, $p=2$, in the absence of charge $Q=0$, and $\alpha=0$ the above equation
reduces up to the first order term of deflection angle for Schwarzschild black hole \cite{A6}
\begin{equation}
\tilde{\alpha}\simeq\frac{4m}{b},
\end{equation}
where $m$ is black hole mass.
The significance of obtained result in weak-field approximation is that the deflection of a light ray is evaluated
by taking domain outside of the lensing region, which implies that the impact of the gravitational
lensing is a global effect in such a way that there are multiple light rays converging
among the source and observer.

Furthermore, by considering the homogeneous plasma medium we evaluate the deflection angle of
photon given by \eqref{sol2} for asymptotically flat hairy BH, which is given below
\begin{eqnarray}
\tilde{\alpha}&=&{\frac {21\,\pi\,{Q}^{2}{p}^{2}{\omega_{e}}^{2}}{64\,{\omega_{\infty}}^{2}
{b}^{2}}}+{\frac {3\,{Q}^{2}{p}^{2}\pi}{32\,{b}^{2}}}+{
\frac {\eta\,{p}^{2}{\omega_{e}}^{2}}{{\omega_{\infty}}^{2}b}}+
{\frac {\eta\,{p}^{2}}{b}}-{\frac {7\,\pi\,\alpha\,{p}^{2}{\omega_{e}}^{2}}{128\,
{\omega_{\infty}}^{2}{b}^{2}{\eta}^{2}}}-{\frac {
\pi\,\alpha\,{p}^{2}}{64\,{b}^{2}{\eta}^{2}}}\nonumber\\
&-&{\frac {{Q}^{2}{\omega_{e}}^{2}}{{\omega_{\infty}}^{2}b\eta}}-{\frac {{Q}^{2}}{b\eta}}+
{\frac {7\,\pi\,{Q}^{2}\alpha\,{\omega_{e}}^{2}}{128\,{\omega_{\infty}}^{2}{\eta}^{4}{b}^{2}}}
+{\frac {\pi\,{Q}^{2}\alpha}{64\,{b}^{2}{\eta}^{4}}}+1/6\,{\frac {\alpha\,
{\omega_{e}}^{2}}{{\omega_{\infty}}^{2}{\eta}^{3}b}}+1/6\,{\frac {\alpha}{b{\eta}^{3}}}.\label{WW2}
\end{eqnarray}

The above equation can be expressed as 
\begin{eqnarray}
\tilde{\alpha}&=&\left[\frac{\eta p^{2}}{b}+\frac{3Q^{2}p^{2}\pi}{32b^{2}}-
\frac{\alpha\pi p^{2}}{64b^{2}\eta^{2}}-\frac{Q^{2}}{b\eta}+
\frac{\alpha Q^{2}\pi}{64b^{2}\eta^{4}}+\frac{\alpha}{6b\eta^{3}}\right]\nonumber\\
&+&\left(\frac{\omega_{e}}{\omega_{\infty}}\right)^{2}
\left[\frac{21\pi Q^{2}p^{2}}{64b^{2}}+\frac{\eta p^{2}}{b}-\frac{7\pi
\alpha p^{2}}{128b^{2}\eta^{2}}-\frac{Q^{2}}{b\eta}
+\frac{7\pi Q^{2}\alpha}{128\eta^{4}b^{2}}+\frac{\alpha}{6\eta^{3}b}\right].
\end{eqnarray}
For $Q=0$ and $p=2$, one can obtain the following form
\begin{eqnarray}
\tilde{\alpha}&=&\left[\frac{4\eta}{b}-\frac{\alpha \pi}{16\eta^{2}b^{2}}+
\frac{\alpha}{6b\eta^{3}}\right]
+\left(\frac{\omega_{e}}{\omega_{\infty}}\right)^{2}\left[\frac{4\eta}{b}-
\frac{7\pi\alpha}{32b^{2}\eta^{2}}+\frac{\alpha}{6\eta^{3}b}\right]\nonumber\\
\tilde{\alpha}&=&\frac{4\eta}{b}\left[1+\left(\frac{\omega_{e}}{\omega_{\infty}}\right)^{2}\right]+
\mathcal{O}(\eta^{2}).
\end{eqnarray}
Due to the presence of plasma medium, the gravitational deflection angle increases,
depends upon the frequency of photons. In homogeneous plasma medium, the photons
having smaller frequency or greater wavelengths, and are deflected by a larger
angle near the gravitating center. For $\omega\rightarrow\omega_{e}$, the
influential difference in the gravitational deflection angles is substantial
for longer wavelengths which is possible only for the radio waves. In this regard, the
gravitational lens in plasma acts as a radio spectrometer \cite{BisnovatyiKogan:2008yg}. Crisnejo and Gallo \cite{A7} have studied the dynamics of light rays in a cold
non-magnetized plasma medium. For this purpose, they have obtained the deflection angle
for Schwarzschild spacetime in homogenous plasma medium
by using Gauss-Bonnet theorem, i.e.,
\begin{equation}
\tilde{\alpha}=\frac{2m}{b}\left(1+\frac{1}{1-(\omega_{e}/\omega_{\infty})^{2}}\right)+
\mathcal{O}(m^{2}).\label{WW3}
\end{equation}
In comparison with the deflection angle obtained in \cite{A7}, with 
the choice of parameters $\eta=m$, $p=2$, $\alpha=0$ and $Q=0$, our proposed deflection angle (\ref{WW2})
approximates the deflection angle of Schwarzschild black hole (\ref{WW3}). In the future, the astrophysical observations might shed light on the effect of hair on deflection angle. Any discovery of the hair would be an important signal beyond the general relativity.

  \acknowledgments
The authors are grateful to anonymous referees for their valuable comments and suggestions to improve the paper. This work is supported by Comisi{\'o}n Nacional de
Ciencias y Tecnolog{\'i}a of Chile (CONICYT) through FONDECYT Grant N{$%
\mathrm{o}$} 3170035 (A. {\"O}.).

\end{document}